\documentclass[useAMS,usenatbib]{mn2e}
\usepackage[totalwidth=515pt,totalheight=680pt,left=1.4cm,right=1.4cm]{geometry}
\usepackage{graphicx,amssymb}
\input psfig

\title[A simple scaling for two-planet instability]
{A simple scaling for the minimum instability time-scale of two widely spaced planets}
\author[Veras \& Mustill]{Dimitri Veras$^{1}$\thanks{E-mail: veras@ast.cam.ac.uk}, Alexander J. Mustill$^{2}$\thanks{E-mail: alex.mustill@uam.es}
\\
$^{1}$Institute of Astronomy, University of Cambridge, Madingley Road, Cambridge CB3 0HA
\\
$^{2}$Universidad Aut\'{o}noma de Madrid, Departamento de F\'{i}sica Te\'{o}rica C-XI, 28049 Madrid, Spain
}

\begin{document}

\date{Accepted 2013 May 20.  Received 2013 May 18; in original form 2013 April 8}

\pagerange{\pageref{firstpage}--\pageref{lastpage}} \pubyear{XXXX} 

\maketitle

\label{firstpage}

\begin{abstract}
Long-term instability in multi-planet exosystems is a crucial consideration when confirming putative candidates, analyzing exoplanet populations, constraining the age of exosystems, and identifying the sources of white dwarf pollution.  Two planets which are Hill stable are separated by a wide-enough distance to ensure that they will never collide.  However, Hill stable planetary systems may eventually manifest Lagrange instability when the outer planet escapes or the inner planet collides with the star.  We show empirically that for two nearly coplanar Hill stable planets with eccentricities less than about 0.3, instability can manifest itself only {\it after} a time corresponding to $x$ initial orbits of the inner planet, where $\log_{10} x \sim 5.2 \left[\mu/(M_{\rm Jupiter}/M_{\odot})\right]^{-0.18}$ and $\mu$ is the planet-star mass ratio.  This relation applies to any type of equal-mass secondaries, and suggests that two low-eccentricity Hill stable terrestrial-mass or smaller-mass planets should be Lagrange stable throughout the main sequence lifetime of any white dwarf progenitor.  However, Hill stable giant planets are not guaranteed to be Lagrange stable, particularly within a few tens of percent beyond the critical Hill separation.  Our scaling represents a useful ``rule of thumb'' for planetary population syntheses or individual systems for which performing detailed long-term integrations is unfeasible. 
\end{abstract}


\section{Introduction}

Stability claims in multi-planet systems are often illusory.  Computational and analytic limitations continue to stymie attempts to fully characterize even just two-planet systems over their entire main sequence lifetimes.  For example, despite highly focused, long-term numerical simulations of the Solar System, the collisional fate of the inner planets remains unknown \citep{batlau2008,lasgas2009}.  \cite{kaietal2011} do integrate the 5-planet 55 Cnc system for over 10 Gyr, but only with subsets of planets, owing to their tight orbits.  Also, despite success at self-consistently integrating two-planet systems with the slow Bulirsch-Stoer algorithm over the entire main sequence lifetime for stars with masses $\ge 3M_{\odot}$ \citep{veretal2013}, nearly all observed exoplanets orbit stars with masses $< 3M_{\odot}$\footnote{See the Extrasolar Planet Encyclopedia at http://exoplanet.eu/.}$^{,}$\footnote{See the Exoplanet Data Explorer at http://exoplanets.org/.}.

We can reduce our reliance on numerical simulations by appealing to special analytical solutions of the three-body problem.  One solution of particular interest to planetary dynamicists defines a boundary beyond which the orbits of two planets can never cross.  This boundary, which was largely broached to the astronomical community by \cite{gladman1993}, has become known as the ``Hill stability boundary", and has been the subject of extensive study (see \citealt*{georgakarakos2008} for a recent review).  However the Hill stable boundary does not indicate whether the inner planet may collide with the star or whether the outer planet can escape the system.  If a system is ``Lagrange stable", then both planets remain bound, never cross orbits, and suffer no collisions.

Despite recent work that explicitly explores the difference between Hill and Lagrange stability \citep{bargre2006,bargre2007,rayetal2009,kopbar2010,veretal2013}, identification of the Lagrange stability boundary in astrophysical contexts remains largely elusive.  No exact analytical solution exists despite much progress in the analysis of the escape of the outer body \citep[Chapter 11 of][]{marchal1990,anosova1996,lietal2009} and the dependence on mass of Lagrange stability \citep{khokuz2011} in Hill stable systems.  Here, we obtain an explicit user-friendly scaling for this dependence.  In doing so, we help provide insight into the location of the fuzzy Lagrange stability separation boundary and hint at its dependence on orbital parameters.

We describe the initial conditions and results of our simulations in Section 2, briefly discuss the results in Section 3 and conclude in Section 4.  

\section{Numerical simulations}

We first must identify the approximate location of the Lagrange stability boundary in order to motivate the construction of initial conditions for identifying a scaling.  Unfortunately, we can never be sure if instability will manifest itself in a system because of the finiteness of the duration of numerical simulations.  

We consider an inner and an outer planet with the same mass $m$, eccentricities $e_1$ and $e_2$, and semimajor axes $a_1$ and $a_2$. The star mass is $M$ such that $\mu \equiv m/M$.  Instability ``occurs'', in the sense that a planet achieves a hyperbolic orbit (``escape'') or collides with the star, at time $t = t_L$, where $t_{L}$ is most naturally measured in number of orbits; we denote $x$ as the number of initial inner planet orbits.  There is a critical ratio $\left(a_2/a_1\right)_{\rm crit}$ beyond which there is a dramatic tail-off of unstable systems.  \cite{bargre2006} found that this ratio resides within just a few percent of the critical Hill stablility ratio for the extrasolar systems 47 Uma and HD 12661.       

Our preliminary simulations suggest that identifying $\left(a_2/a_1\right)_{\rm crit}$ becomes more difficult for higher values of $e_1$ and $e_2$, when the boundary is ``leakier'' (chaotically diffuse).  The likely reason is that the range of potential initial velocities increases with higher $e$, and hence the initial locations on the osculating planetary orbits become more important.  These velocities are intrinsically linked to Lagrange stability \citep[e.g.][]{sos2005}.  Hence, we restrict our analysis to low eccentricities in order to achieve a robust result. Our goal is to obtain a function for $t_L$ or $x$ as a function of $\mu$.

\begin{figure*}
\centerline{\LARGE \bf Instability Times in log$\mathbf{_{10}{(x)}}$ vs. Initial Semimajor Axis Ratio}
\centerline{
\psfig{figure=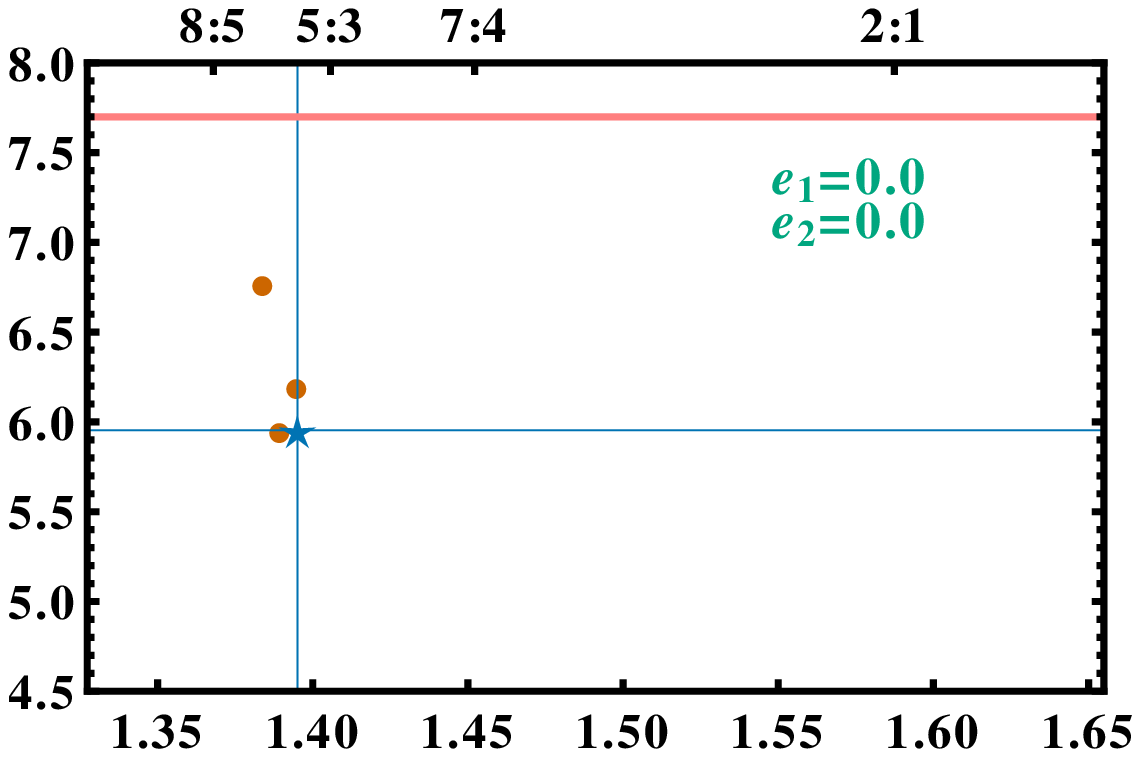,height=4.0cm,width=4.4cm}
\psfig{figure=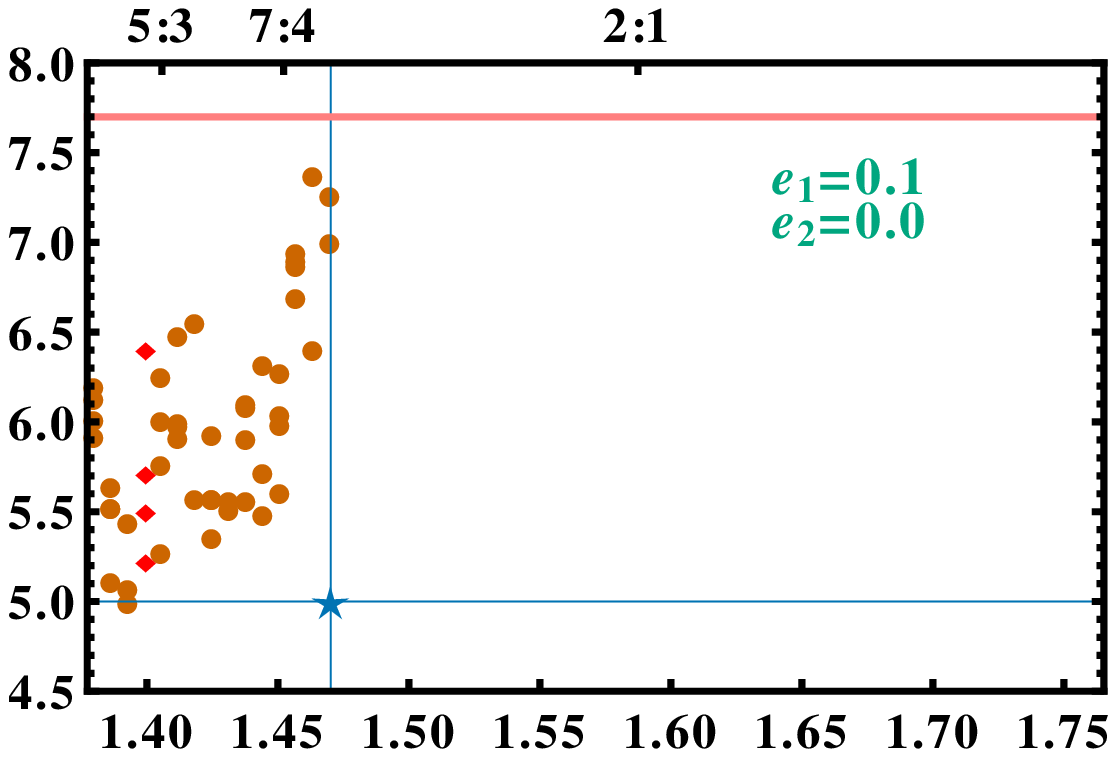,height=4.0cm,width=4.4cm}
\psfig{figure=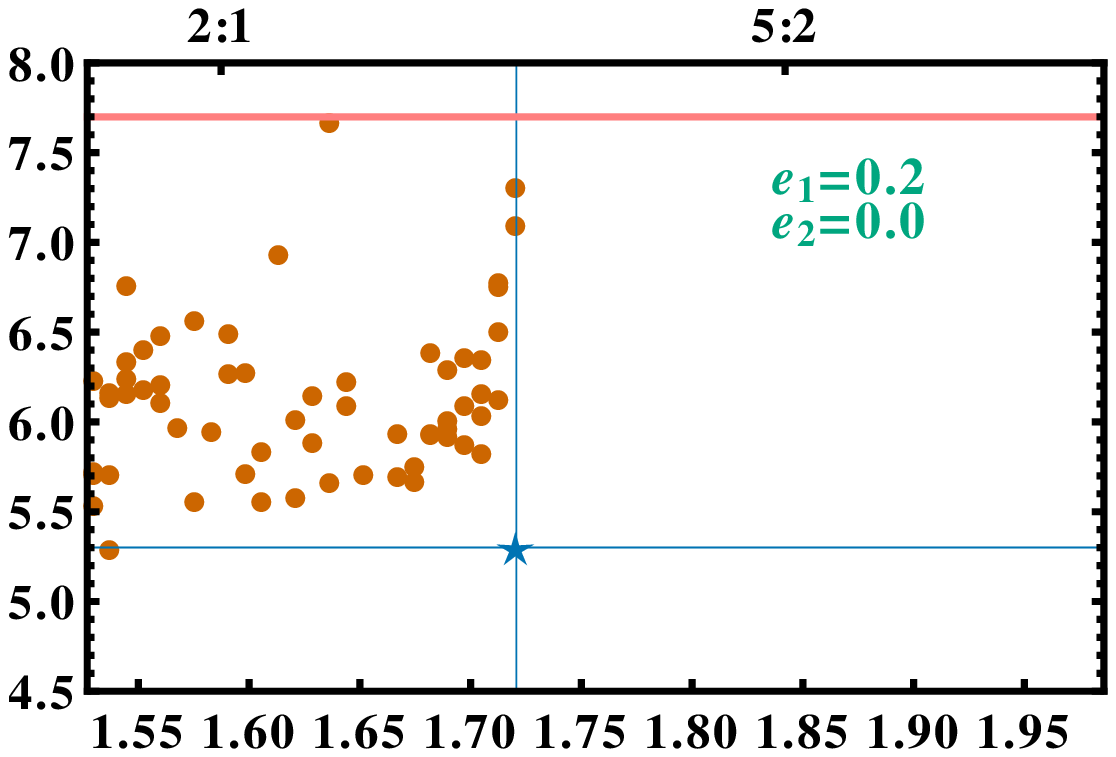,height=4.0cm,width=4.4cm}
\psfig{figure=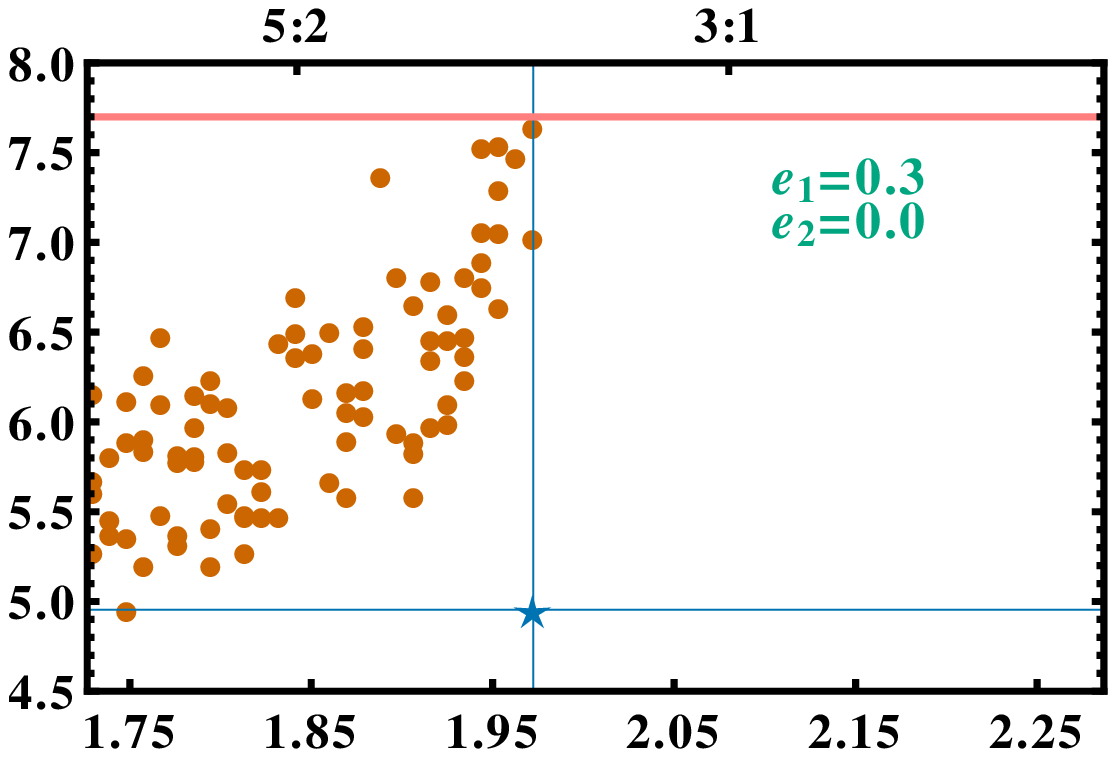,height=4.0cm,width=4.4cm}
}
\vspace{-20pt}
\centerline{
\psfig{figure=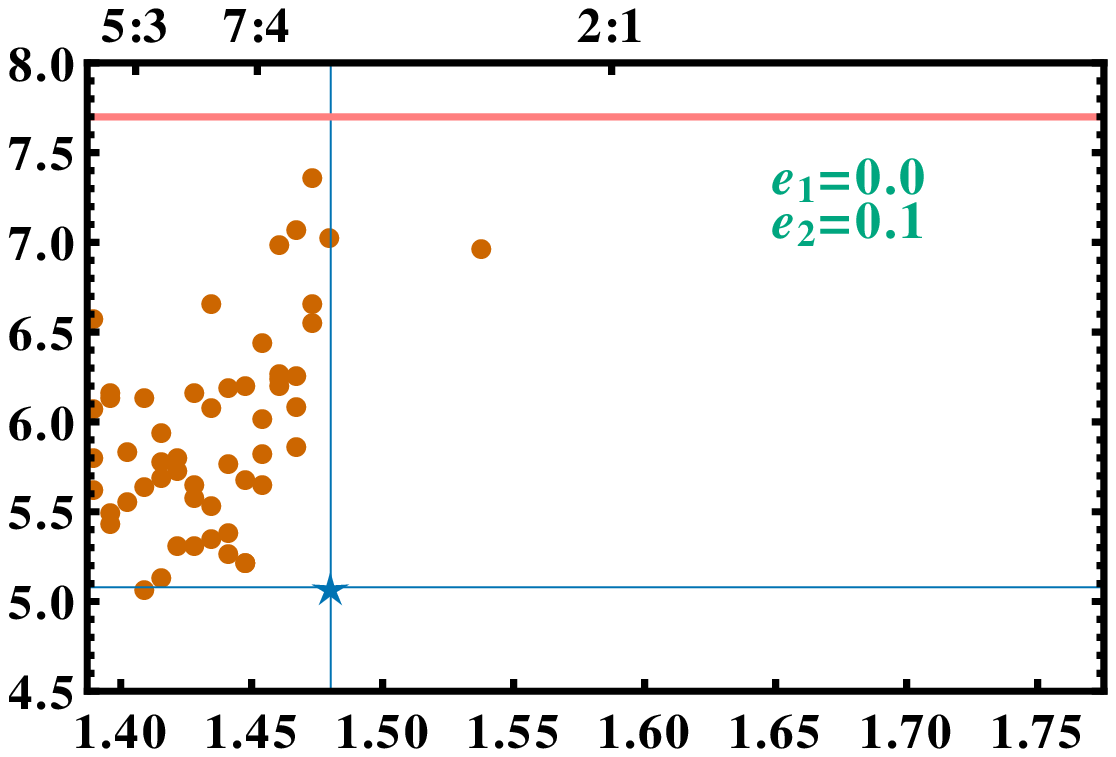,height=4.0cm,width=4.4cm}
\psfig{figure=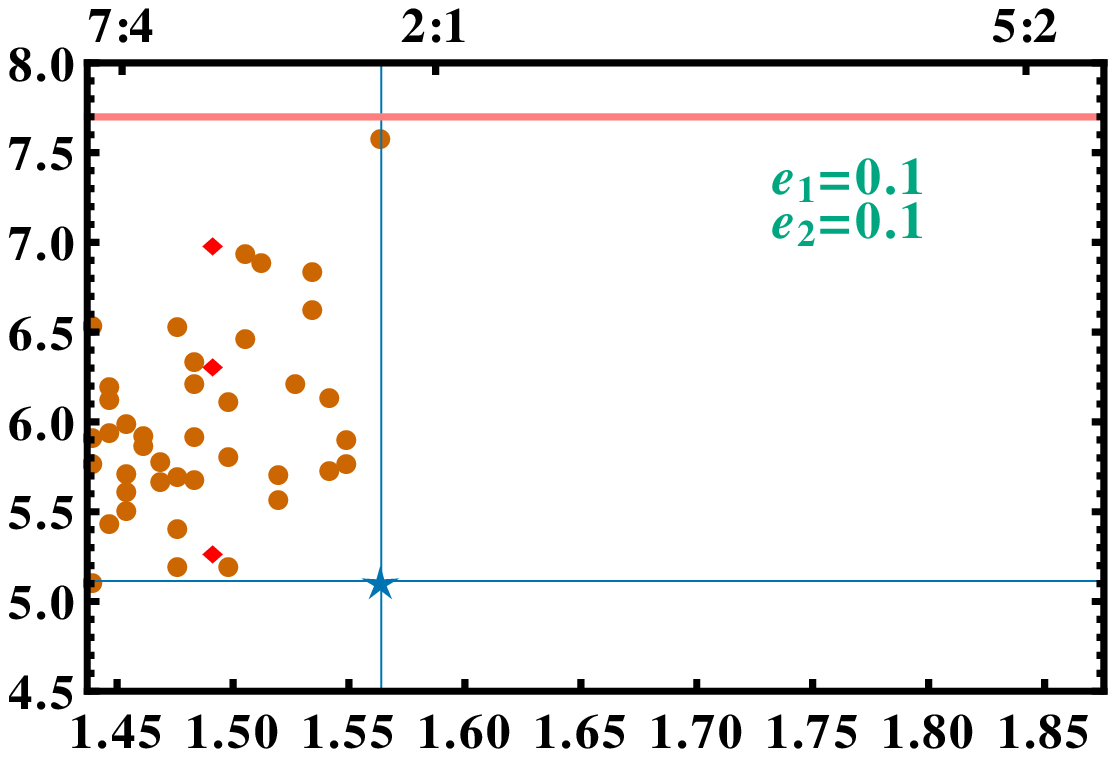,height=4.0cm,width=4.4cm}
\psfig{figure=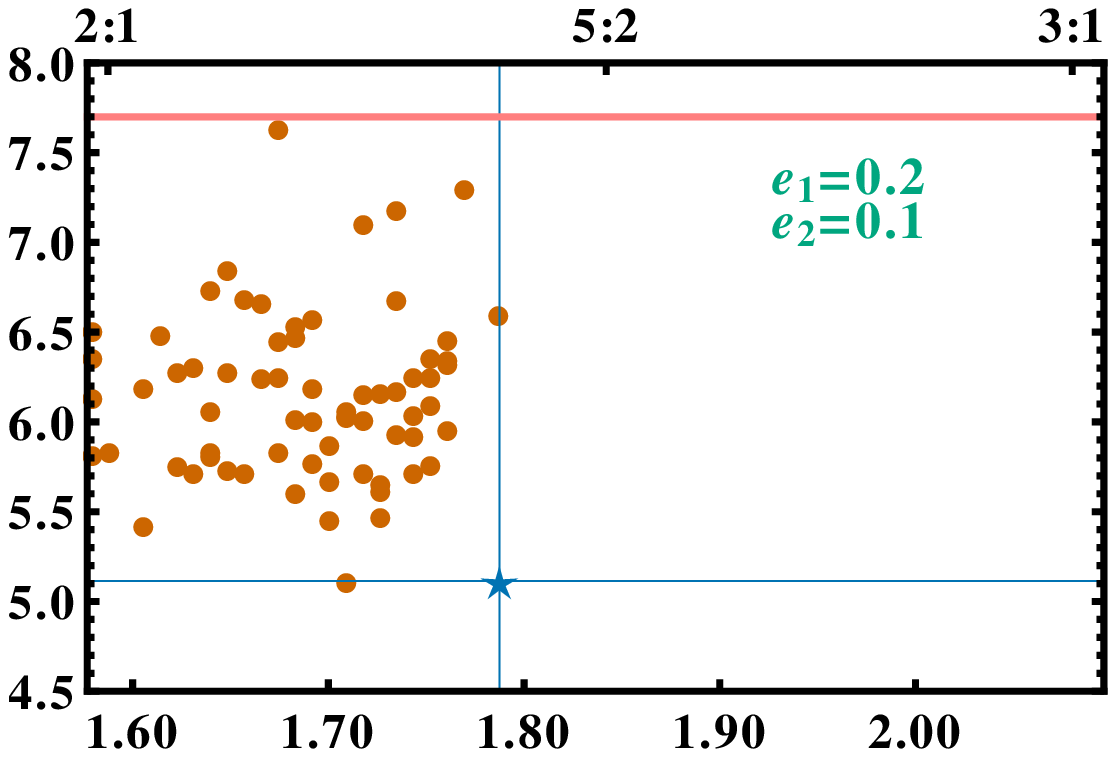,height=4.0cm,width=4.4cm}
\psfig{figure=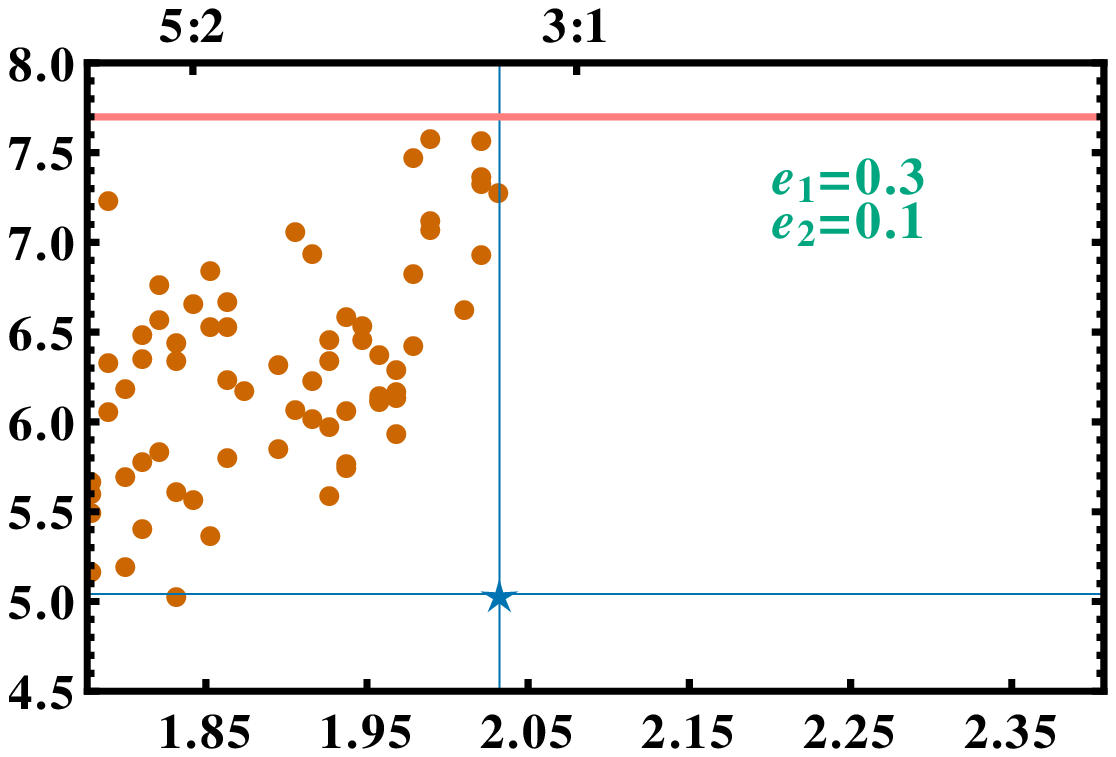,height=4.0cm,width=4.4cm}
}
\vspace{-20pt}
\centerline{
\psfig{figure=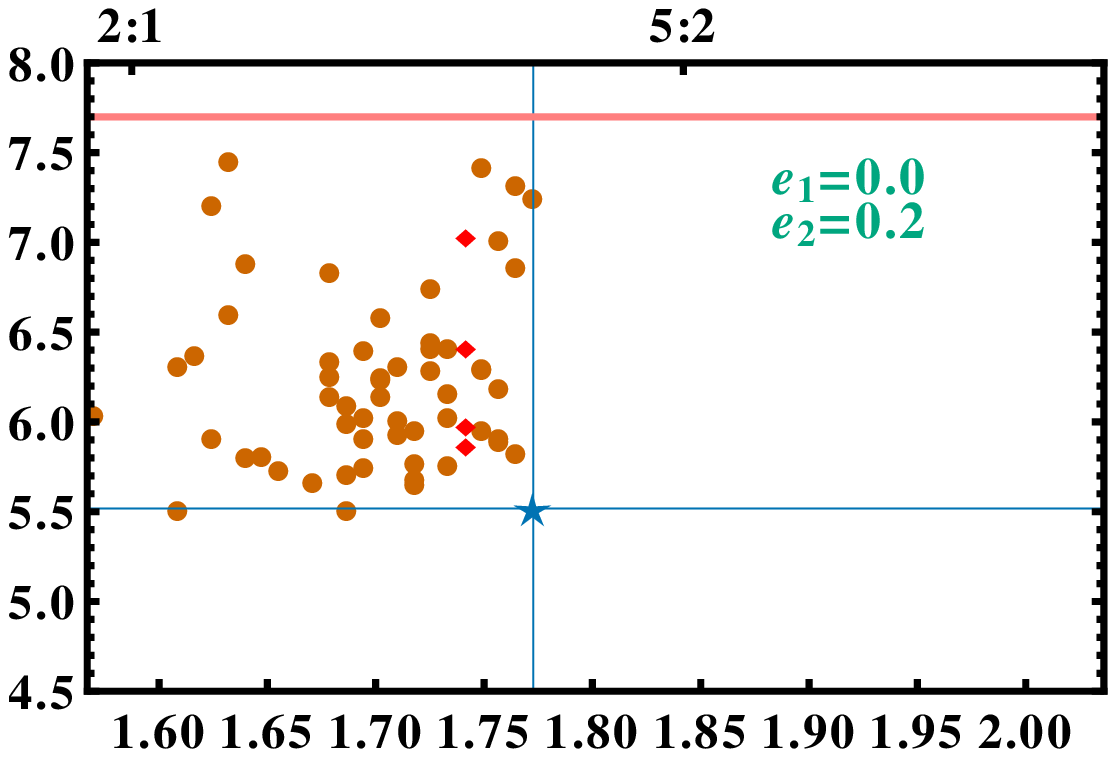,height=4.0cm,width=4.4cm}
\psfig{figure=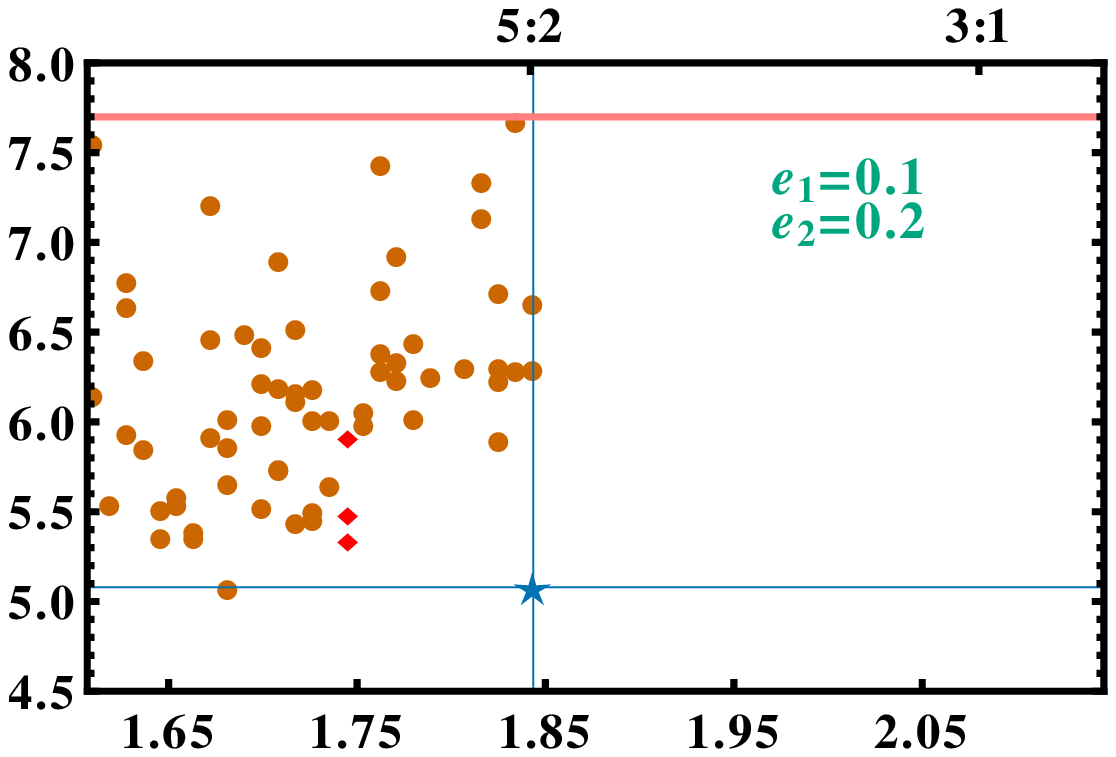,height=4.0cm,width=4.4cm}
\psfig{figure=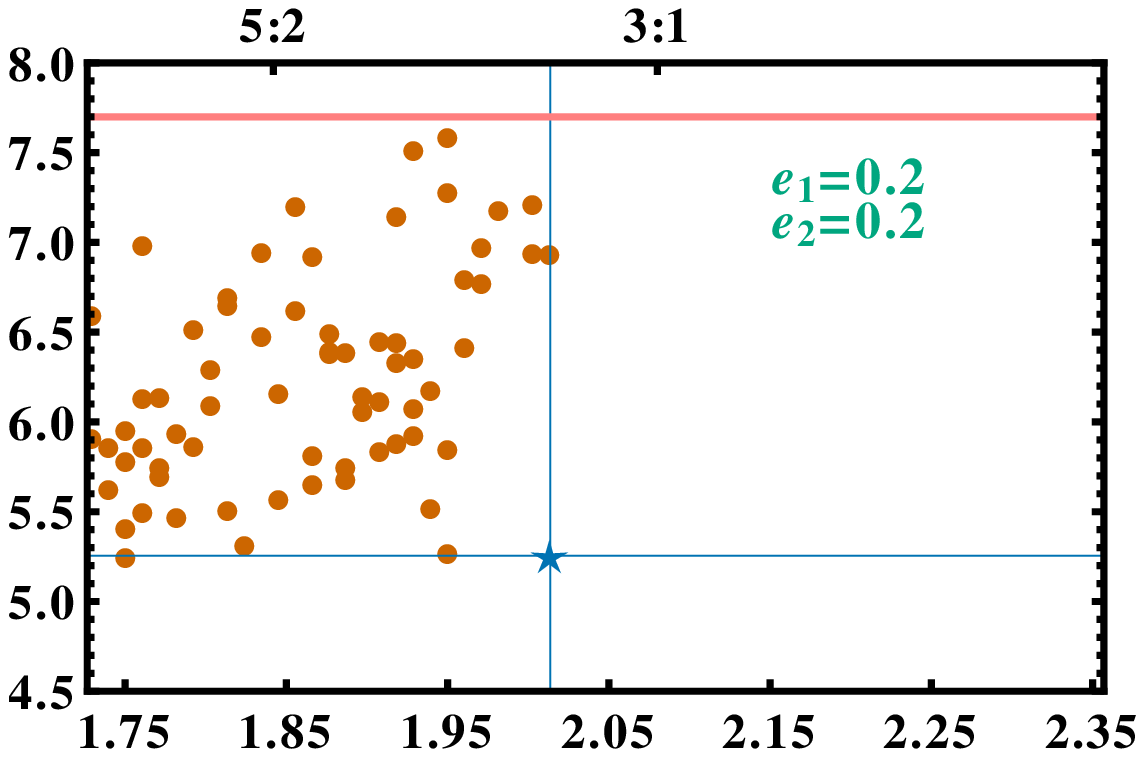,height=4.0cm,width=4.4cm}
\psfig{figure=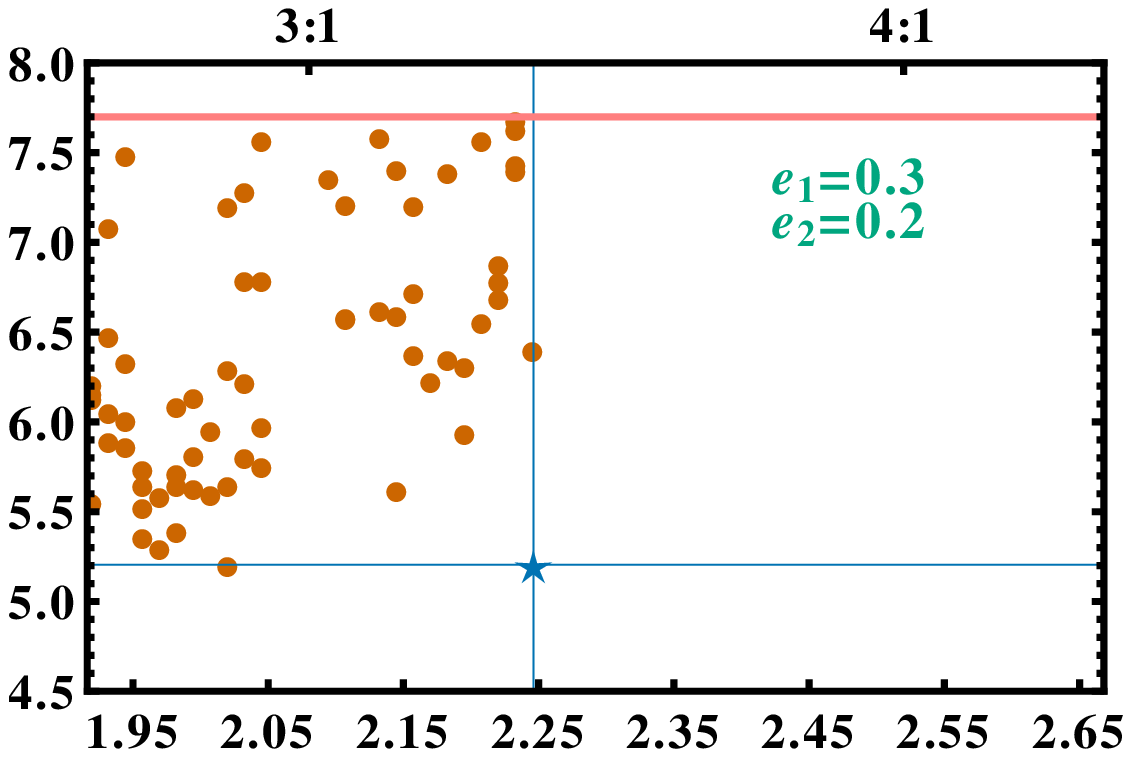,height=4.0cm,width=4.4cm}
}
\vspace{-20pt}
\centerline{
\psfig{figure=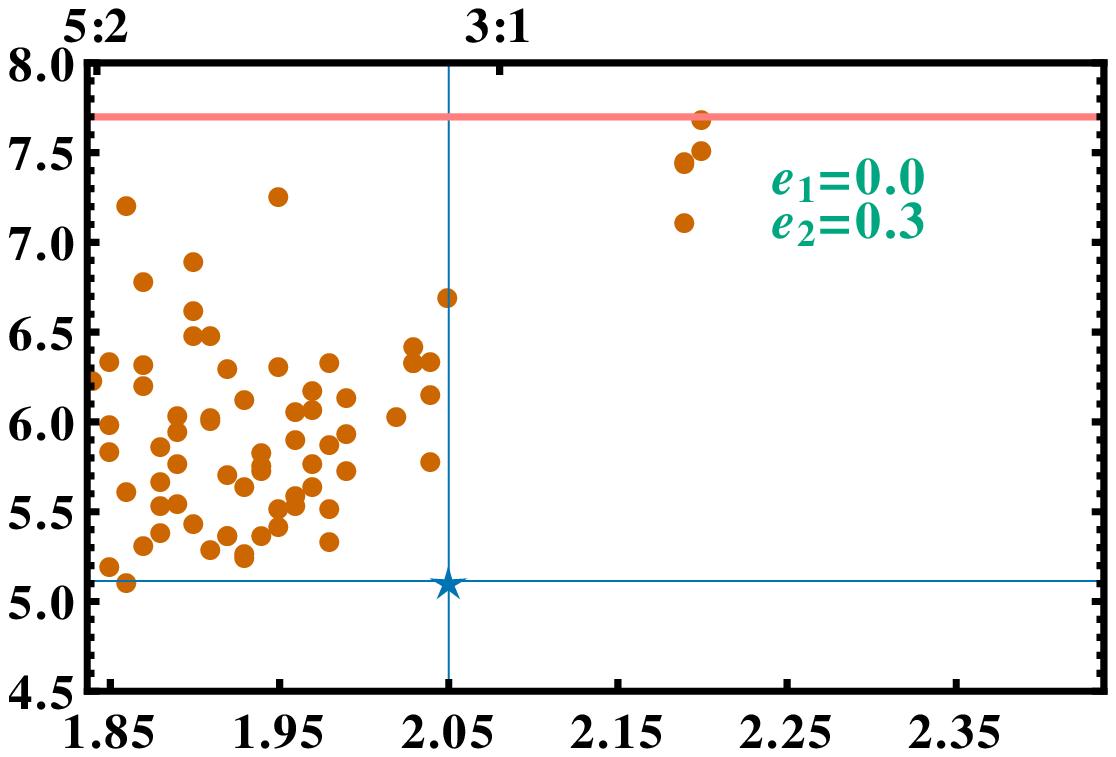,height=4.0cm,width=4.4cm}
\psfig{figure=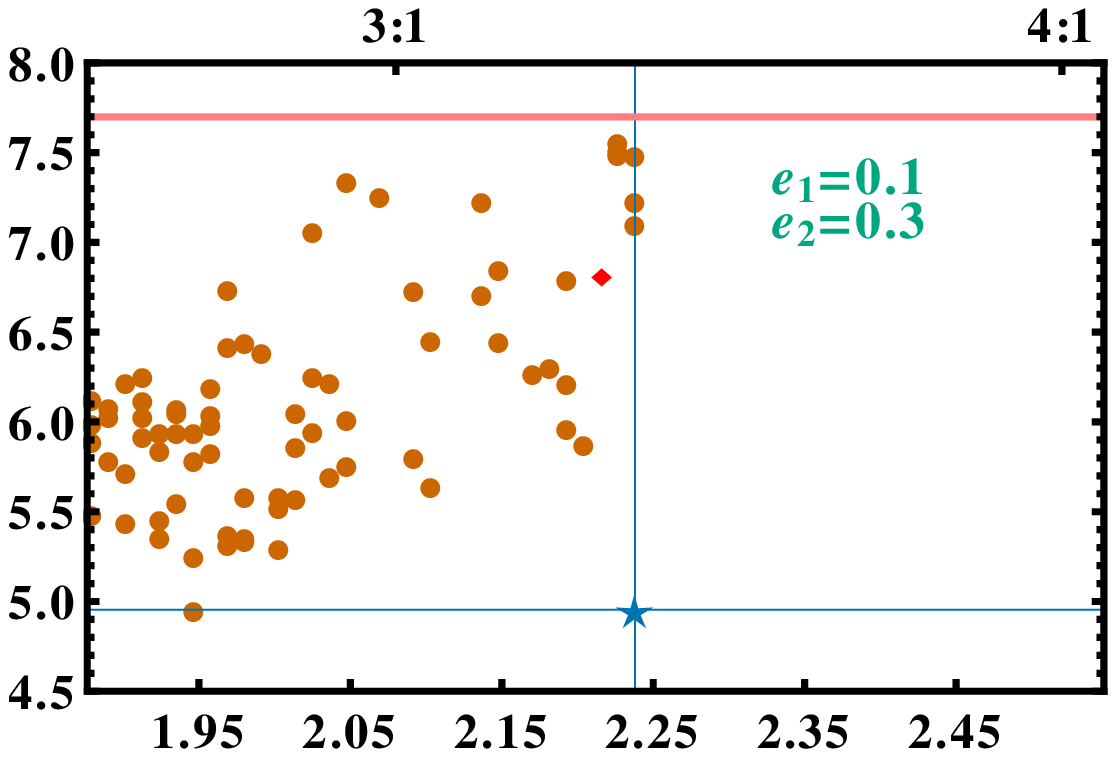,height=4.0cm,width=4.4cm}
\psfig{figure=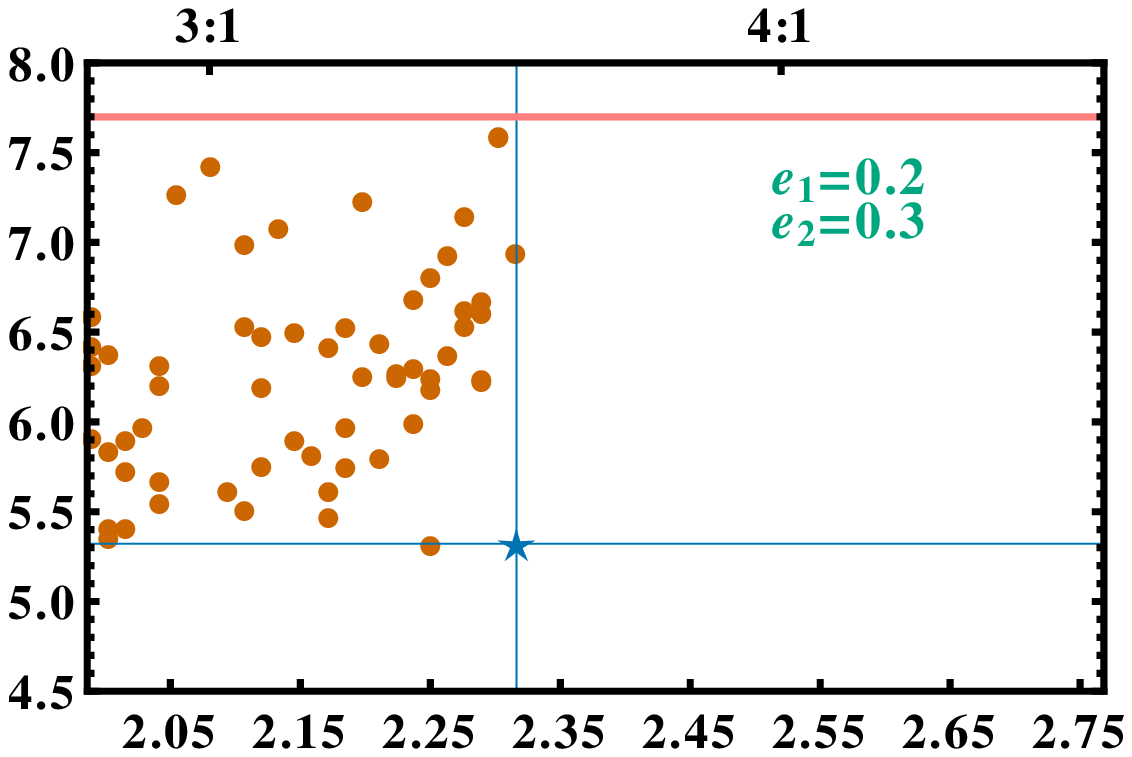,height=4.0cm,width=4.4cm}
\psfig{figure=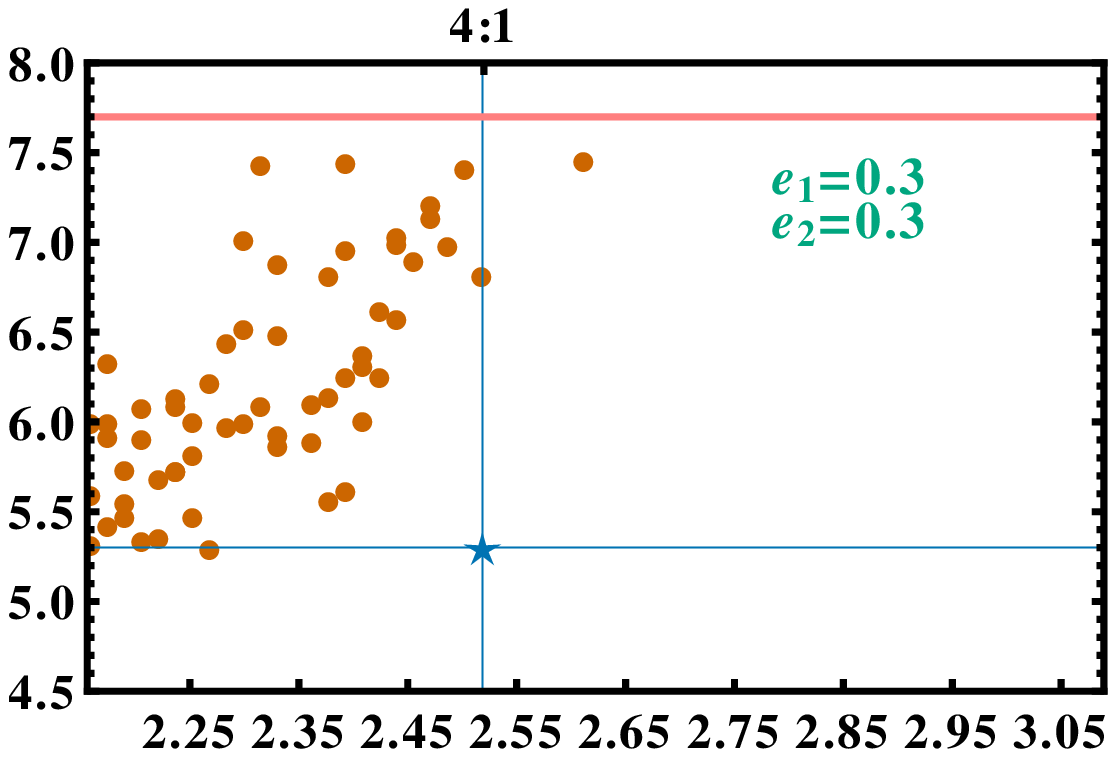,height=4.0cm,width=4.4cm}
}
\caption{
Bounding instability for low-eccentricity 
Hill-stable $m = 1 M_J$ planets.  Each panel represents a different
combination of initial eccentricities.  The value of the
left y-axis in all plots represents the critical Hill separation,
rounded up to the nearest hundredth in semimajor axis ratio.
Orange dots refer to systems that feature outer 
planet escape, whereas red diamonds illustrate that
the inner planet collided with the star.  The horizontal
blue lines bound the instability by providing the minimum
instability time.  The vertical blue lines are arbitrarily
drawn at the largest initial 
separation within which are included 
the vast majority of unstable systems over
$5 \times 10^7$ inner planet orbits.  
The plot primarily demonstrates that the minimum 
instability times are largely independent of initial
eccentricity.
}
\label{figs1MJ}
\end{figure*}

\begin{figure*}
\centerline{
\psfig{figure=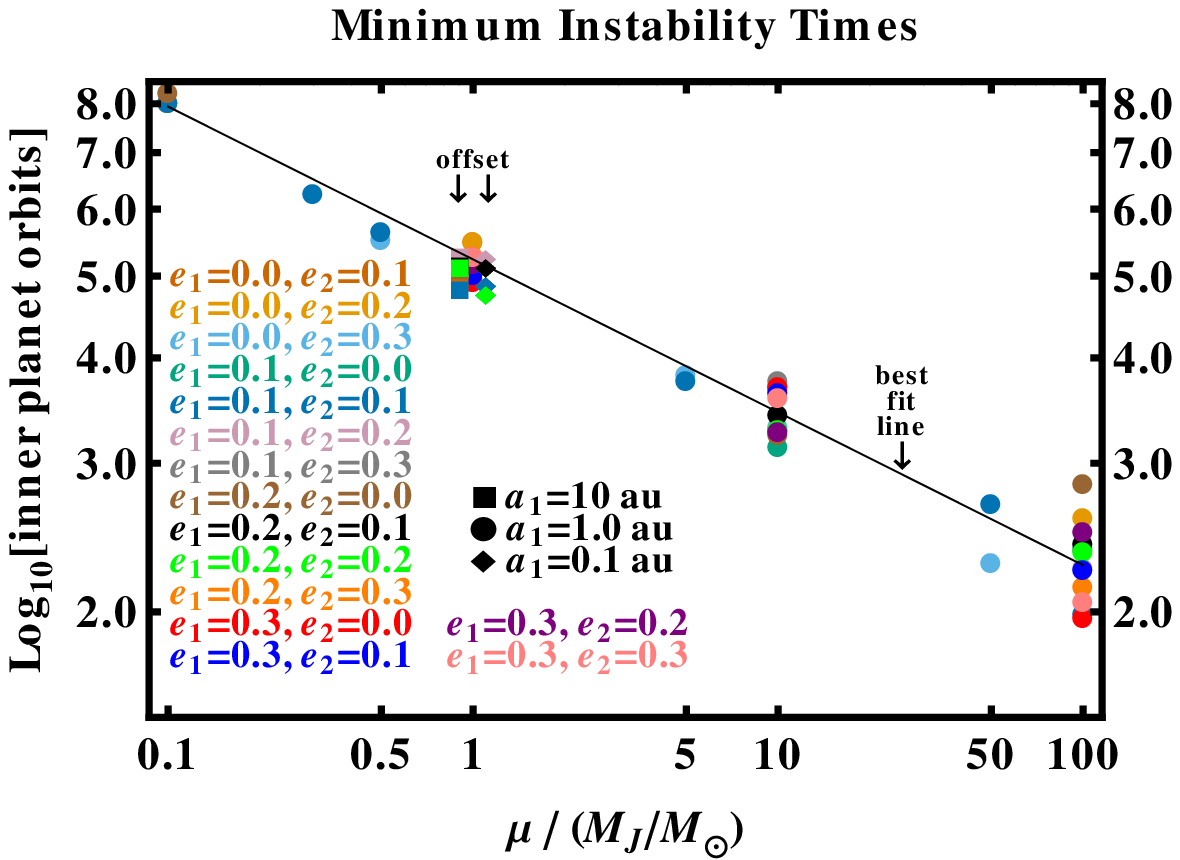,height=8.0cm,width=13.0cm} 
}
\caption{
Minimum instability times as a function of 
the secondary/primary mass ratio for two equal-mass Hill-stable 
secondaries.  Simulations for $a_1 = 0.1$ au and $a_1 = 10$ au were computed
for $\mu = 1M_{J}/M_{\odot}$, but are shown slightly offset for
clarity.  The main result of this paper, a fit to the 
straight-line trend of the circle symbols apparent in 
this figure, is given in equation 1.
}
\label{Mplot}
\end{figure*}

\subsection{Numerical integrator}

To carry out the simulations, we use the conservative Bulirsch-Stoer integration routine (bs2) from a modified version of the {\tt MERCURY} integration package \citep{chambers1999}.  The modification is in the collision detection subroutine {\tt mce\_cent} and is described in Section 3.3 of \cite{veretal2013}.  This modification is necessary for accurately simulating very massive (up to about $100 M_{J}$) secondaries which may achieve highly eccentric orbits.  We use a tolerance parameter of $10^{-12}$, which yields typical pre-instability energy and angular momentum errors of less than $10^{-4}$ and $10^{-6}$ respectively, even for our longest simulations.

\subsection{Preliminary Considerations}

We consider only Hill-stable systems. We compute the Hill stability boundary according to the procedure in \cite{veretal2013}, which utilizes formulae from both \cite{donnison2006} and \cite{donnison2011}.  These formulae, which are re-expressed versions of the general equations from \cite{marboz1982}, allow the Hill boundary to be computed implicitly in terms of arbitrary eccentricities and inclinations, and the masses of the three bodies, all expressed in Jacobi coordinates.  Allowing for arbitrary eccentricities and utilizing Jacobi coordinates is crucial, particularly for high $\mu$ where application of astrocentric coordinates would blur the stability boundaries.  Our treatment of Hill stability is more general than the typically-used criterion from \cite{gladman1993} partly because his equation 14 restricts the secondary masses to be much smaller than the primary mass.

Hill stability ensures that planet-planet collisions never occur in our simulations.  However, planet-star collisions do occur, and we can compare the frequency of these collisions with the frequency of systems featuring planetary escape. This comparison will be of interest to, for example, investigators seeking a possible dynamical trigger \citep{bonetal2011,debetal2012,veretal2013} for the accretion processes which cause abundant metal pollution in white dwarf atmospheres \citep[e.g.][]{zucetal2003} and likely accompany the associated dusty debris discs \citep[e.g.][]{faretal2009} and gaseous debris discs \citep[e.g.][]{ganetal2008}. Furthermore, planets scattered onto orbits with small pericentres may become tidally circularised before colliding with the star, which has been proposed as an origin of the hot Jupiter population \citep{forras2006}.

We sample systems from the Hill stability boundary out to a semimajor axis ratio which well exceeds $\left(a_2/a_1\right)_{\rm crit}$.  Because we do not know where this critical boundary lies {\it a priori}, we perform broad preliminary sweeps of phase space to help focus our more detailed simulations.

Also, the duration of our simulations must exceed $t_L$.  Figure 9 of \cite{veretal2013} suggests that for $a_1 = 10$ au, $m = 1M_J$, $e_1 = e_2 = 0.1$ and parent star masses of $M_{\star} = \lbrace 5 M_{\odot}, 4 M_{\odot}, 3 M_{\odot} \rbrace$, then $t_L \approx \lbrace 7 \times 10^7 {\rm yr}, 5 \times 10^7 {\rm yr}, 2 \times 10^7 {\rm yr}\rbrace$.  Hence, in these cases instability occurs after $x \approx \lbrace 5 \times 10^6, 3 \times 10^6, 1 \times 10^6 \rbrace$ (orbits of the inner planet) for planet-star mass ratios of $\mu \approx \lbrace 1.9 \times 10^{-4}, 2.4 \times 10^{-4}, 3.2 \times 10^{-4} \rbrace$.  These values provide useful guides for our initial conditions.

\subsection{Fiducial simulations}

Let us choose a higher mass ratio of $\mu = 1M_{J}/1M_{\odot} \approx 9.54 \times 10^{-4}$ and integrate our systems for much longer, for $x = 5 \times 10^7$.  We also sample mass ratios one and two orders of magnitude higher, so that we can characterize instabilities at shorter times.  We set $M_{\star} = 1M_{\odot}$ and the stellar radius to be one Solar radius in all cases.  The entire integration is assumed to take place on the main sequence.  We also set $a_1 = 1$ au unless otherwise indicated.

We choose the other parameters as follows.  We sample 4 different initial values of eccentricity ($\left\lbrace 0.0, 0.1, 0.2, 0.3 \right\rbrace$) for both $e_1$ and $e_2$.  We assume the two planets are nearly coplanar, and assign them each a random inclination up to $0.01^{\circ}$.  We extract their longitudes of ascending nodes from a uniform random distribution, as we do their mean anomalies and arguments of pericentres.  Therefore, the maximum initial mutual inclination between the planets is $0.02^{\circ}$.

In each set of simulations, we sample 60 different values of $\left(a_2/a_1\right)$ uniformly from the Hill stability limit to a value equal to Hill limit multiplied by a factor of $\left[1 + k(1+e_1)(1+e_2)\right]$, where $k = 0.25$ or $k = 0.30$ for $m = 1 M_J$ or $m > 1 M_J$ respectively.  These values were chosen based on extensive preliminary simulations and in order to best showcase our result.  For each value of $\left(a_2/a_1\right)$, we simulate 4 different systems, each with randomized orbital angles.

\subsubsection{Results}

The result for our $\mu = 1M_{J}/1M_{\odot}$ simulations is presented in Fig. \ref{figs1MJ}.  Although simulations were run over the entire horizontal extent of each plot, only unstable systems are marked with orange dots (escape) or red diamonds (planet-star collision).  The horizontal blue line indicates the minimum instability time, and the vertical blue line provides a rough estimate of $\left(a_2/a_1\right)_{\rm crit}$.  The pink horizontal line indicates the duration of the simulation ($x = 5\times 10^7$).  The upper axes display all 1st- to 3rd-order mean motion commensurable locations present for the semimajor axes ratio sampled.  Noteworthy observations include

\begin{itemize}
\item{The minimum instability time occurs at $10^5 \lesssim x \lesssim 10^{5.5}$ in each plot except for the $e_1 = e_2 = 0$ case, where nearly every Hill stable system is Lagrange stable.}
\item{Instability occurs in a continuous manner for increasing $\left(a_2/a_1\right)$ until $\left(a_2/a_1\right)_{\rm crit}$.  However, in three cases ($e_1=0.0,e_2=0.1$ ; $e_1=0.0,e_2=0.3$ ; $e_1=e_2=0.3$), a few instances of instability occur well beyond this boundary.  If the cause is proximity to mean motion commensurabilities, they must be weak (higher than 3rd-order).}
\item{The value of $\left(a_2/a_1\right)_{\rm crit}$ does not appear to be robust for the highest eccentricity cases, where $e_1 = 0.3$ or $e_2 = 0.3$, because the critical value seems to be limited by the integration duration.}
\item{The predominant outcome of instability exhibited is planetary escape.  The occurrence of collisions seems to be a strong function of $\left(a_2/a_1\right)$, as evidenced by the red diamonds appearing in vertical strips in the ($e_1=0.1,e_2=0.0$ ; $e_1=e_2=0.1$ ; $e_1=0.0,e_2=0.2$ ; $e_1=0.1,e_2=0.2$) cases.}
\item{The existence of the minimum instability time-scale does not guarantee that systems will become unstable at or close to that value: instability may be delayed beyond $100t_L$.}
\end{itemize}

Outcomes of the simulations for $\mu = 10M_{J}/1M_{\odot}$ and $\mu = 100M_{J}/1M_{\odot}$ also exhibit the trends stated in the last four above bullets.  However, the minimum instability times of these simulations are distinctly different, and suggest a simple scaling with $\mu$.

\subsection{Additional simulations}

To help establish our scaling result, we obtain more minimum instability times with additional 
simulations for $m = 50M_J, 5M_J, 0.5M_J, 0.3M_J$, and $0.1M_J$, and for $a_1 = 0.1$ au and $10$ au.  We 
need to consider different values of $a_1$ because unlike Hill stability, Lagrange stability will not 
necessarily be scalable with $a_1$. The reason is a stellar collision requires the planet to hit a 
target whose solid angular size decreases with the square of the distance, and escape requires the outer 
planet to have sufficient energy or momentum to leave the influence of the star's gravitational potential.

For each simulation set, we ran a total of 80 simulations, sampling 4 orbital configurations for each of 20 
values of $\left(a_2/a_1\right)$ very close to the Hill limit (within a tenth in semimajor axis ratio, 
evenly spaced) to help ensure that instability will occur.  For the lowest-mass cases, we extended the duration of
some simulation sets until one simulation featured instability.  For the highest-mass cases,
we shortened the interval between outputs to obtain representative values.

\subsubsection{Results}

We combined the results of these simulations with those of our fiducial simulations to produce 
Fig. \ref{Mplot}.  The different symbols on the figure refer to different values of $a_1$; for clarity,
the squares and diamonds are offset from the circles for $\mu = 1M_J/M_{\odot}$ by $0.1 M_J/M_{\odot}$ in each
direction.  The minimum instability time appears to be insensitive to the choice of $a_1$.    
A least-squares fit to all of the $a_1 = 1.0$ au points on the plot suggests the following scaling:

\begin{equation}
\log_{10}{x} \sim 5.2 \left(\frac{\mu}{M_J/M_{\odot}}\right)^{-0.18}
.
\end{equation}

Regarding the meaning of $x$, the duration of one orbit is derived from the value of $a_1$, which 
is the initial value of the inner semimajor axis.  The value of $a_1$ may change with time if 
the planets become locked in a mean motion
resonance.  Regardless, the usefulness of equation 1 relies on the relation being measured 
with respect to initial parameters.

We emphasize that this scaling is approximate, but nevertheless provides a reliable
order-of-magnitude lower bound for instability timescales over at least 4 orders of magnitude
in mass ratio.  The scaling also satisfies the perfectly circular case, which is not plotted
because in some of those simulation ensembles all systems remained stable.

\begin{figure}
\centerline{
\psfig{figure=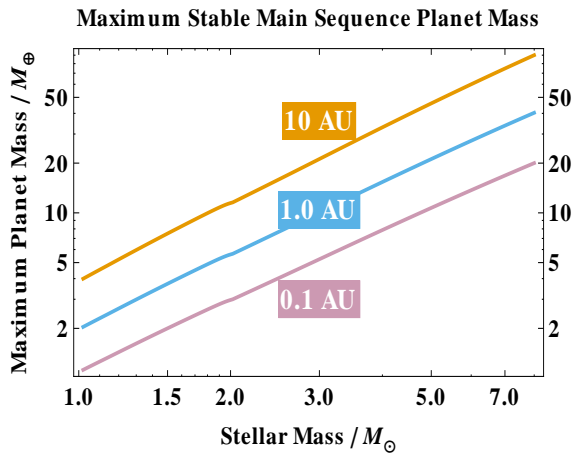,height=6.0cm,width=8.0cm} 
}
\caption{
Maximum planet mass for which an equal mass two-planet Hill-stable system
will remain Lagrange stable (bound and ordered) throughout the entire main
sequence lifetime.  Jovian-mass planets are always susceptible
to instability for $a < 10$ AU.
}
\label{MS}
\end{figure}

\section{Discussion}

We may extrapolate the scaling to determine the maximum planet-star mass ratio
for which a Hill-stable two-planet system is guaranteed to remain Lagrange stable
over the entire main sequence lifetime of a star.  We compute main sequence 
lifetimes from the {\tt SSE} code \citep{huretal2000}.  Figure \ref{MS} plots the
result, and demonstrates that Hill stable terrestrial-mass planets are always Lagrange stable
for $a > 0.1$ au, whilst Hill-stable Jovian-mass planets are never guaranteed to be Lagrange
stable for $a < 10$ au.  The fate of Super-Earths would depend on the details of the system
in question.

\cite{bargre2006} suggest that Hill stability may be used as a proxy for
stability over the main sequence.  Our results suggest that this correlation
is robust for orbiting bodies which are less massive than a 
terrestrial-mass.  Their tentative conclusion that two planets with a semimajor
axis ratio larger than about 120 per cent of the Hill stability limit is Lagrange stable
is consistent with the parameter space explored in our Fig. \ref{figs1MJ}.  However, the 
effect of external perturbations, such as from star-planet tides or 
other planets in the same system, remain unclear, and a potential future avenue of study.

Three other potential extensions may include (1) testing the robustness of equation
1 for highly eccentric planets and significantly inclined planets, (2)
developing an analytic explanation for the dependencies in equation 1,
and (3) better defining $\left(a_2/a_1\right)_{\rm crit}$.
Our preliminary simulations suggest that equation 1 remains valid for highly
eccentric planets, whereas $\left(a_2/a_1\right)_{\rm crit}$ breaks down quickly
as $e$ increases beyond 0.3.  The scaling in equation 1 is not unique given 
the scatter in the points in Fig. \ref{Mplot}, but was chosen for simplicity. 
In order to motivate analytical studies, we note that the exponent of $-0.18$ is very close
to $-2/11 \approx -0.181$, and could easily instead be fit to 
$-1/5 = -0.200$ or $-1/6 \approx -0.167$ for different coefficients and different sets
of model simulations.

Finally, Fig. \ref{figs1MJ} hints that for a given value of $\left(a_2/a_1\right)$, there may exist a {\it maximum} number of orbits at which a system could manifest Lagrange instability.  This possibility is best visualized for the $e_2 = 0.0$ cases.  This boundary is likely to be diffuse, but may provide a useful metric to help describe system stability.  Unfortunately, computational limitations continue to inhibit such explorations.

\section{Conclusion}

Characterizing the stability of exoplanetary systems has become 
a vital component of discovery papers and population synthesis models.
However, no analytical formulation exists for the Lagrange stability boundary of a 
two-planet system in time nor space.  Here, we have empirically identified this 
boundary in time (equation 1), finding a minimum time-scale for 
instability as a function of planet mass, and demonstrated that the boundary is independent
of semimajor axis and eccentricity at least for $e_1,e_2 < 0.3$ and $0.1$~au~$\lesssim a_1 \lesssim 10$~au.  
The implications are that 
Hill stability may be equated with Lagrange stability only for objects that
are less massive than roughly $1M_{\oplus}$, whereas Hill-stable Jovian-mass
planets are always susceptible to eventual instability along the main sequence.

\section*{Acknowledgments}

We thank the referee, Richard Greenberg, for his useful perspective and suggestions for 
clarifications, resulting in an improved manuscript.  We also thank 
Sverre J. Aarseth for fruitful discussions.  This work made use of facilities funded by the European Union 
through ERC grant number 279973. AJM is supported by Spanish grant AYA 2010/20630.

\label{lastpage}
\end{document}